\documentclass[aps,pre,superscriptaddress,showpacs,twocolumn]{revtex4}

\usepackage{graphicx}
\usepackage{amssymb}
\usepackage{amsmath}
\usepackage{amsfonts}
\usepackage{mathrsfs}
\usepackage{color}

\begin{document}

\title{Quantitative analysis of single particle trajectories:
mean maximal excursion method}

\author{Vincent Tejedor}
\affiliation{Physics Department, Technical University of Munich, James
Franck Strasse, 85747 Garching, Germany}
\author{Olivier B{\'e}nichou}
\affiliation{Laboratoire de Physique Th{\'e}orique de la Mati\`ere
Condens\'ee, Universit{\'e} Pierre et Marie Curie, Paris, France}
\author{Raphael Voituriez}
\affiliation{Laboratoire de Physique Th{\'e}orique de la Mati\`ere
Condens\'ee, Universit{\'e} Pierre et Marie Curie, Paris, France}
\author{Ralf Jungmann}
\affiliation{Physics Department, Technical University of Munich, James
Franck Strasse, 85747 Garching, Germany}
\author{Friedrich Simmel}
\affiliation{Physics Department, Technical University of Munich, James
Franck Strasse, 85747 Garching, Germany}
\author{Christine Selhuber-Unkel}
\affiliation{Niels Bohr Institute, Blegdamsvej 17, 2100 K{\o}benhavn,
Denmark}
\author{Lene B. Oddershede}
\affiliation{Niels Bohr Institute, Blegdamsvej 17, 2100 K{\o}benhavn,
Denmark}
\author{Ralf Metzler}
\affiliation{Physics Department, Technical University of Munich, James
Franck Strasse, 85747 Garching, Germany}

\begin{abstract}
An increasing number of experimental studies employ single particle tracking
to probe the physical environment in complex systems. We here propose and
discuss new methods to analyze the time series of the particle traces,
in particular, for subdiffusion phenomena.
We discuss the statistical properties of mean maximal excursions,
i.e., the maximal distance covered by a test particle up to time $t$.
Compared to traditional methods focusing on the mean squared displacement
we show that the mean maximal excursion analysis performs better in the
determination of the anomalous diffusion exponent. We also
demonstrate that combination of regular moments with moments of the mean
maximal excursion method provides additional criteria to determine the exact
physical nature of the underlying stochastic subdiffusion
processes. We put the methods to test using
experimental data as well as simulated time series from different models
for normal and anomalous dynamics, such as diffusion on fractals,
continuous time random walks, and fractional Brownian motion.\\

\end{abstract}


\maketitle

\section{Introduction}

The history of stochastic motion may be traced back to the writings of Titus
Lucretius, describing the battling of dust particles in air \cite{lucretius}.
Later, irregular motion of single coal dust particles was described by
Jan Ingenhousz in 1785 \cite{ingenhousz}. Robert Brown in 1827 reported the
jittery motion of small particles within the vacuoles of pollen grains
\cite{brown}. Possibly the first systematic recording of actual trajectories
was published by Jean Perrin, observing individual, small granules in
uniform gamboge emulsions \cite{perrin}. Yet apparently the first
experimental study based on the time series analysis of single
particle trajectories is due to Nordlund who tracked small mercury
spheres in water \cite{nordlund}.
Today single trajectory analysis is a common method to probe the motion of
particles, notably, in complex biological environments
\cite{seisenhuber,golding,elbaum,elbaum1,lene,lene1,platani,pan,wong,garini}.

Typically a diffusion process in $d$ dimensions is characterized by the
ensemble averaged mean squared displacement (MSD)
\begin{equation}
\label{eamsd}
\langle\mathbf{r}^2(t)\rangle=\int_0^{\infty}r^2P(r,t)dV=2dK_{\alpha}t^{\alpha}.
\end{equation}
Here we assumed spherical symmetry and an isotropic environment, such that
$P(r,t)$ is the probability density to find the particle a (radial)
distance $r$ away from the origin at time $t$ after release of the
particle at $r=0$ at time $t=0$. In equation (\ref{eamsd}) we introduced
the anomalous diffusion exponent $\alpha$. In the limit $\alpha=1$
we encounter regular Brownian diffusion. For other values of $\alpha$ the
associated diffusion is anomalous: the case $0<\alpha<1$ is called
subdiffusion while for $\alpha>1$ the process is superdiffusive
\cite{report}. In this work we focus on subdiffusive processes. In
equation (\ref{eamsd}) the generalized diffusion coefficient is of
dimension $[K_{\alpha}]=\mathrm{cm}^2 /\mathrm{sec}^{\alpha}$. Subdiffusion
of the form (\ref{eamsd}) is found in a variety of systems, such as
amorphous semiconductors \cite{scher}, tracer dispersion in subsurface
acquifers \cite{grl}, or in turbulent systems \cite{paolo}.

In fact, subdiffusion was found from observation of single trajectories in
a number of biologically relevant systems: For instance, it was shown that
adeno-associated viruses of radius $\approx 15$ nm in a cell perform
subdiffusion with $\alpha=0.5\ldots0.9$ \cite{seisenhuber}. Fluorescently
labeled messenger RNA chains of 3000 bases length and effective diameter of
some 50nm subdiffuse with $\alpha\approx0.75$ \cite{golding}. Lipid granules
of typical size of few hundred nm exhibit subdiffusion with
$\alpha\approx0.75\ldots0.85$ \cite{elbaum,elbaum1,lene,lene1}; and the
diffusion of telomeres in the nucleus of mammalian cells shows
$\alpha\approx0.3$ at shorter times, and $\alpha \approx0.5$ at
intermediate times \cite{garini}. A study assuming normal diffusion for
the analysis of tracking data of single cell nuclear organelles shows
extreme fluctuations of the diffusivity as function of time along
individual trajectories, possibly
pointing to subdiffusion effects \cite{platani}. \emph{In vitro,}
subdiffusion was measured in protein solutions \cite{pan} and in
reconstituted actin networks \cite{wong}. Molecular crowding is often
suspected as a cause of subdiffusion in living cells 
\cite{Minton,Saxton}.

Currently one of the important open questions is what physical mechanism
causes the subdiffusion in biological systems. Single particle tracking is
expected to provide essential clues to answer this question. Thus, recently
a method has been suggested based on the statistics of first passage times,
i.e., the distribution of times it takes a random walker to first reach a
given distance from its starting point. This quantity has been shown to be
a powerful tool to discriminate between CTRW and diffusion on fractals
\cite{PNAS,Olive07}. However such an analysis requires a huge amount of
data to be statistically relevant.
Fluorescence correlation spectroscopy (FCS) has also been proposed to
identify the physical mechanism of subdiffusion \cite{weiss}; but this
approach is based on an indirect observable, the fluorescence correlator,
which is not directly comparable with analytical results; moreover this
method needs to fit three parameters to a single curve. We here present a
new method, that is based on analytical results. Our approach is
demonstrated to enable one to extract more, and more accurate, information
from a set of single particle trajectories.

A typical single particle tracking experiment provides a time series
$\mathbf{r}(t)$ of the particle position from which one may calculate the
time averaged mean squared displacement
\begin{equation}
\label{tamsd}
\overline{\delta^2(\Delta,T)}=\frac{1}{T-\Delta}\int_0^{T-\Delta}\Big[
\mathbf{r}(t+\Delta)-\mathbf{r}(t)\Big]^2dt.
\end{equation}
Here $T$ denotes the overall measurement time, and $\Delta$ is a lag time
defining a window swept over the time series. For a Brownian random walk
with typical width $\langle\delta\mathbf{r}^2\rangle$ of the step length
and characteristic waiting time $\tau$ between successive steps, we recover
the time average $\left<\overline{\delta^2(\Delta,T)}\right>=2dK_1\Delta$,
where the diffusion constant becomes $K_1=\langle\delta\mathbf{r}^2\rangle/
[2 d \tau]$. In this case the time average provides exactly the same
information as the ensemble average. Note that this is not always the case
when the dynamics is subdiffusive \cite{he,lubelski,deng}.

Using time averages to analyze the behavior of a single particle is an
elegant method, in particular, to avoid errors from averages over particles
with nonidentical physical properties. However in many cases the actual
trajectories are too short to allow one to extract meaningful information
from the time average. Moreover, in cases where the
subdiffusion is governed by a CTRW with diverging characteristic waiting
time the values of the moments, and therefore their ratios, become random
quantities \cite{he,lubelski}. Using the ensemble average prevents this
problem. We
therefore consider herein ensemble averages calculated directly from
measured trajectories.
In particular we present an analysis based on a mean maximal excursion
statistics. It will be shown that this method provides relevant
information on the system, complementary to results from analysis of
regular moments. Moreover we demonstrate that the mean maximal excursion
method may obtain more accurate information about the dynamics than the
typically measured mean squared displacement (\ref{eamsd}).

In what follows we present the theoretical background of the mean maximal 
excursion analysis and discuss how different dynamic processes can be
discriminated. We then discuss how to apply these methods in practice,
including the analysis of some recent single particle tracking data.

\section{Materials and Methods}
\label{Experimental data}

As a benchmark for our quantitative analysis we here define the three most
prominent approaches to subdiffusion. Physically these processes
are fundamentally different, while they all share the form (\ref{eamsd})
of the mean squared displacement. In the supplementary material
we provide details on how we simulate the time series based on the
stochastic models.

\emph{(i) Continuous Time Random Walk (CTRW).}
CTRW defines a random walk process during which the walker rests a random
waiting time, drawn from a probability distribution, between successive
steps \cite{scher}. If the density of waiting times is of the long tailed
form
\begin{equation}
\label{wtd}
\psi(t)\sim \frac{\alpha\tau^{\alpha}}{\Gamma(1-\alpha)t^{1+\alpha}},
\end{equation}
for $0<\alpha<1$, the mean waiting time $\int_0^{\infty}t\psi(t)dt$
diverges, and the resulting process becomes subdiffusive with mean squared
displacement (\ref{eamsd}). The exponent $\alpha$ from the waiting time
density (\ref{wtd}) is then the same as in equation (\ref{eamsd}). If the
variance of the associated jump lengths is again
$\langle\delta\mathbf{r}^2\rangle$, the generalized diffusion coefficient
becomes $K_{\alpha}=\langle\delta\mathbf{r}^2\rangle/(2 d \tau)$. Waiting
times with such power-law distribution were, for instance, observed for
the motion of probes in a reconstituted actin network \cite{wong}. CTRW is
used in a wide variety of fields, ranging from charge carrier motion in
amorphous semiconductors \cite{scher}, over tracer diffusion in
underground aquifers \cite{grl}, up to weakly chaotic systems
\cite{paolo}.

\emph{(ii) Diffusion on fractals.}
A random walker moving on a geometric fractal, for instance, a percolation
cluster near the percolation threshold, meets bottlenecks and dead ends on
all scales, similar to the motion in a labyrinth. This results in an
effective subdiffusion in the embedding space. While the fractal dimension
$d_f$ characterizes the geometry of the fractal, the diffusive dynamics
involves an additional critical exponent, the random walk exponent $d_w$
($d_w \ge 2$). The latter is related to the anomalous diffusion exponent
through $\alpha = 2/d_w$ \cite{havlin}.
Fractals can be used to model complex networks, and have recently been
suggested to mimic certain features of diffusion under conditions of
molecular crowding \cite{yossi,olivier}.
We will use for the theoretical descriptions the dynamical scheme of
reference \cite{Procaccia}.

\emph{(iii) Fractional Brownian Motion (FBM).}
FBM was introduced to take into account correlations in a random walk: the
state of the system at time $t$ is influenced by the state at time $t'<t$.
In the FBM model this is achieved by passing from a Gaussian white noise
$dB(t)$ to fractional Gaussian noise
\begin{eqnarray}
\nonumber
B_H(t)&=&\frac{1}{\Gamma(H+1/2)}\left(\int_0^t(t-\tau)^{H-1/2}dB(\tau)\right.\\
&&\hspace*{-1.2cm}+\left.
\int_{-\infty}^0\Big[(t-\tau)^{H-1/2}-(-\tau)^{H-1/2}\Big]dB(\tau)\right),
\label{fbmnoise}
\end{eqnarray}
where the Hurst exponent $0<H<1$ is connected to the anomalous diffusion
exponent by $\alpha=2H$. FBM therefore describes both subdiffusion and
superdiffusion up to the ballistic limit $\alpha=2$. FBM is used to
describe the motion of a monomer in a polymer chain \cite{gleb} or single
file diffusion \cite{tobias}. FBM has recently been proposed to underlie
the diffusion in a crowded environment \cite{weiss}. The autocorrelation
function of FBM in 1D reads \cite{mandelbrot}
\begin{equation}
\langle X^H(t_1)X^H(t_2)\rangle=\frac{K_1}{2}\left(t_1^{2H}+t_2^{2H}-
|t_1-t_2|^{2H}\right)
\end{equation}
and for $t_1=t_2$ we recover the mean squared displacement (\ref{eamsd}).
Following reference \cite{unterberger}, we extend FBM to several dimensions
such that a $d$-dimensional FBM of exponent $H$ is a process in which each
of the coordinates follows a one-dimensional FBM of exponent $H$. The
resulting $d$-dimensional FBM still satisfies (\ref{eamsd}), with
$\alpha=2H$.

\section{Results}
\label{theory}

The parameters in the three simulation models are chosen to produce the same
anomalous diffusion exponent $\alpha=0.70$. Using only the classical
analysis based on the MSD (\ref{eamsd}), one could not tell which model
was used to create the data. We discuss here how additional observables
allow one to extract a more accurate value of this $\alpha$ exponent, and
how they may be used to distinguish the microscopic stochastic mechanisms.

\subsection{Mean maximal excursion (MME) approach}
\label{alpha}

A power law fit to the classical MSD (\ref{eamsd}) provides the magnitude
of the anomalous diffusion exponent $\alpha$. We here show that the MME
method is a better observable to determine $\alpha$. The maximal excursion
is the greatest distance $r$, that the random walker reaches until time
$t$. This quantity is averaged over all trajectories, to obtain the MME
second moment
\begin{equation}
\label{mme2}
\langle r_{\mathrm{max}}^2 (t)\rangle=\int_0^{\infty}r_0^2\mathrm{Pr}\left(
r_{\mathrm{max}}=r_0, t\right)dr_0,
\end{equation}
where $\mathrm{Pr}\left(r_{\mbox{\scriptsize max}}=r_0,t\right)$ is the
probability that the maximal distance from the origin that is reached up to
time $t$, is equal to $r_0$. The MME second moment (\ref{mme2}) scales
like $t^{\alpha}$, as shown in reference \cite{Chave} for fractal media,
and derived in the supplementary material for a CTRW process.

For FBM this quantity is not known, similar to the first passage in other
than a semi-infinite domain in 1D. However, one can still use the MME
method to numerically analyze data created by an FBM process, as shown
below.

Why is the MME second moment better than the more standard MSD? The ratio
$\gamma=\sigma_X(t)/\langle X(t)\rangle$ of the standard deviation $\sigma_X
(t)=\sqrt{\langle(X(t)-\langle X(t)\rangle)^2\rangle}$ versus the mean is a
measure for the dispersion around the center of the distribution (first
moment). A lower ratio means that the random variable has a smaller spread
around its mean. This will produce a smoother average and thus a more
accurate fit as the larger number of data points closer to the average
value receive a higher relative weight. Indeed, for regular Brownian
motion the ratio is smaller for the MME second moment than for the regular
second moment, the time independent values being
$\gamma(\mathrm{MSD})/\gamma(\mathrm{MME})=1.61,$ 1.44, and 1.34 for one,
two, and three dimensions.
The MME method is therefore expected to non-negligibly outperform the MSD
method. Details of this calculation are presented in the supplementary
material. For diffusion on a fractal, the ratio
$\gamma(\mathrm{MSD})/\gamma(\mathrm{MME})$ also grows with decreasing
fractal dimension, being always greater than $1$.
For a CTRW the ratio $\gamma(\mathrm{MSD})/\gamma(\mathrm{MME})$
diminishes as well with decreasing $\alpha$, reaching its lowest value at
$\alpha=0$. But it is always larger than 1 in dimensions $d=1,2,3$.

Another way to characterize the dispersion of the MME method versus regular
moments is the ratio of the fourth moment versus the second moment of the
respective distribution:
(i) For a random walk on a fractal, approximated by the dynamical scheme of
reference \cite{Procaccia}, the MME moments become \cite{Chave}
\begin{equation}
\langle r_{\mathrm{max}}^k\rangle=A_{k,d_f,\alpha}\left(\frac{K}{\alpha^2}t
\right)^{k\alpha/2},
\label{Proc-MME}
\end{equation}
where the prefactor is given through
\begin{equation}
A_{k,d_f,\alpha}=\frac{2^{1-\alpha d_f/2}k\alpha}{\Gamma(k\alpha/2+1)\Gamma
(\alpha d_f/2)}\int_0^{\infty}\frac{u^{\alpha(2k+d_f)/2-2}}{I_{\alpha
d_f/2-1}(u)}du.
\label{A-MME}
\end{equation}
Here $I_n$ is the modified Bessel function of the first kind. The regular
moments satisfy an analogous relation \cite{Procaccia},
\begin{equation}
\langle r^k\rangle=\frac{\Gamma(\alpha[k+d_f]/2)}{\Gamma(\alpha
d_f/2)}\left(4Kt\right/\alpha^2)^{k\alpha/2}.
\label{Proc-MSD}
\end{equation}
The ratios $\langle r_{\mathrm{max}}^4\rangle/\langle r_{\mathrm{max}}^2
\rangle^2$ and $\langle r^4\rangle/\langle r^2\rangle^2$ are therefore time
independent numerical constants. Note that above expressions also contain
the limiting case of Brownian motion (integer dimension, and $\alpha=1$). In
the latter case the associated values are listed in table \ref{rtab},
demonstrating again that the MME distribution is more concentrated and
therefore more amenable to parameter extraction by fitting, see also the
discussion below.

\begin{table}
\centering
\begin{tabular*}{\hsize}{@{\extracolsep{\fill}}ccccc}
 & $\alpha$ & 1 D & 2 D & 3 D \cr
 \hline
$\langle r^4 \rangle/\langle r^2 \rangle^2$ & 1 & 3 & 2 & 5/3 \cr
$\langle r_{\mathrm{max}}^4\rangle/\langle r_{\mathrm{max}}^2\rangle^2$ & &
1.77 & 1.49 & 1.36 \cr\hline
$\langle r^4 \rangle/\langle r^2 \rangle^2$ & 1/2 & $3\pi/2\approx4.71$ &
$\pi\approx3.14$ & $5\pi/6\approx2.62$ \cr
$\langle r_{\mathrm{max}}^4\rangle/\langle r_{\mathrm{max}}^2\rangle^2$ & &
2.78 & 2.33 & 2.14 \cr
\hline
\end{tabular*}
\caption{Ratios of fourth moment versus the square of the second moment for
normal moment statistics and MME statistics. We list normal Brownian motion
($\alpha=1$) and CTRW subdiffusion with $\alpha=1/2$. The MME distribution
is narrower and therefore more amenable for data fitting in all cases.}
\label{rtab}
\end{table}

(ii) For FBM, the regular moments are obtained from the Brownian ones by
simple replacement of time $t$ by $t^{\alpha}$. Since the regular moment
ratios are time independent we find exactly the same values as in
the Brownian case. The MME moments are not known analytically, so we
performed numerical simulations to get an estimate of these quantities. A
surprising result is that the MME moments $\langle
r_{\mathrm{max}}^k\rangle$ are proportional to $t^{k\alpha'/2}$, but with
a new exponent $\alpha'>\alpha$.

We discuss these results in detail in the supplementary material, finding a
linear correlation ($R^2>0.999$ for 10 points) between the two exponents:
\begin{equation}
\alpha'\approx0.156\pm0.005+(0.849\pm0.008)\alpha
\label{eq:FBM-MME}
\end{equation}
We note that for Brownian motion ($\alpha = 1$), we retrieve the classical
result $\alpha'=\alpha$. We also obtained an expression for the MME moment
ratio, $\langle r_{\mathrm{max}}^4\rangle/\langle r_{\mathrm{max}}^2
\rangle^2$, in 2D ($R^2 > 0.99$ for 10 points):
\begin{equation}
\frac{\langle r_{\mathrm{max}}^4\rangle}{\langle r_{\mathrm{max}}^2
\rangle^2}\approx(1.05\pm0.01)\left(\frac{\alpha}{2}\right)^{1.42\pm0.01}
+(1.10\pm0.01).
\label{eq:FBM-ratio}
\end{equation}
We note that solely focusing on the determination of $\alpha'$ from the
second MME moment may lead to an overestimation of the anomalous diffusion
exponent if the motion is governed by FBM and $\alpha'$ is not converted to
$\alpha$ via relation (\ref{eq:FBM-ratio}). It is therefore important to also
evaluate the complementary criteria such as the mean squared displacement and
the moment ratios.

(iii) In the case of CTRW subdiffusion we profit from the fact that in
Laplace space we can transform the probability density and the moments of
normal Brownian motion into the corresponding CTRW subdiffusion solution
by
so-called subordination \cite{feller,report}. In practice this means that we
can replace $s$ by $K_1s^{\alpha}/K_{\alpha}$ where $s$ is the Laplace
variable conjugated to time $t$. We obtain the ratio for both regular
moments and MME statistics from the Brownian result, however, with
different pre-factors
\begin{eqnarray}
\langle r^k\rangle_{\mathrm{CTRW}}&=&\frac{\Gamma(k/2+1)}{\Gamma(\alpha
k/2+1)} \langle r^k\rangle_{\mathrm{BM}}, \label{CTRW-MSD}\\
\langle
r^k_{\mathrm{max}}\rangle_{\mathrm{CTRW}}&=&\frac{\Gamma(k/2+1)}{\Gamma(\alpha
k/2+1)} \langle r^k_{\mathrm{max}}\rangle_{\mathrm{BM}}.
\label{CTRW-MME}
\end{eqnarray}
Table \ref{rtab} shows the results for $\alpha=1/2$.

The moment ratios $\langle r_{\mathrm{max}}^4\rangle/\langle
r_{\mathrm{max}}^2 \rangle^2$ and $\langle r^4\rangle/\langle
r^2\rangle^2$ are useful observables. Once we determine the anomalous
exponent $\alpha$ from fit to the MSD or the second MME moment we can use
the moment ratios to identify the process. If the moment ratio for a
subdiffusion process with $0<\alpha<1$ is the same as for Brownian motion
we are dealing with
an FBM process. If the value matches the one for CTRW subdiffusion for the
given $\alpha$ we verify the CTRW mechanism. Finally, we can identify the
remaining possibility, i.e., diffusion on a fractal: The obtained numerical
value for the ratio allows us, in principle, to deduce the underlying
fractal dimension $d_f$, using the predicted values of equation
(\ref{Proc-MME}) and (\ref{Proc-MSD}). We will discuss below how
reliable such classifications are.

\subsection{Determination of the fractal dimension $d_f$}

Finally we establish a criterion to distinguish diffusion on a fractal
from CTRW and FBM subdiffusion. We know that the probability density for a
diffusing particle on a fractal satisfies the scaling relation
\cite{Havlin,metz}
\begin{equation}
P(r,t)=t^{-\alpha d_f/2}P\left(\frac{r}{t^{\alpha/2}},1\right).
\end{equation}
The same relation holds for a CTRW or a FBM if we replace $d_f$ by the
Euclidian dimension. Let us focus on the probability to be in
a growing sphere of radius $r_0 t^{\alpha/2}$. Then
\begin{eqnarray}
\nonumber
\mathrm{Pr}\left(r\leq r_0t^{\alpha/2},t\right)&=&\int_0^{r_0t^{\alpha/2}}
r^{d-1}P(r,t)dr\\
&=&A(r_0)t^{\alpha(d-d_f)/2}.
\end{eqnarray}
Since the exponent $\alpha$ is known from the second MME moment fit we can
extract $d_f$ from above relation.

\subsection{Summary}

Collecting the results from this section we come up with the following
recipe to analyze diffusion data obtained from experiment or simulation,
compare also the results summarized in table \ref{sumtab}.

\begin{table*}
\begin{tabular} {c|c|c|c}
 & Second moment (regular, MME) & Ratio (regular, MME) & Growing spheres\\
\hline
BM & ($\propto t$,$\propto t$) & $(2,1.49)$, eq. (\ref{Proc-MME}) and
(\ref{Proc-MSD}) & $\mathrm{Pr}\left(r\leq r_0t^{\alpha/2},t\right) = A_0$\\
\hline
Fractals & ($\propto t^{\alpha}$,$\propto t^{\alpha}$) & $(<2,<1.49)$, eq.
(\ref{Proc-MME}) and (\ref{Proc-MSD}) & $\mathrm{Pr}\left(r\leq
r_0t^{\alpha/2},t\right) \propto t^{\alpha (2-d_f)/2}$\\
\hline
CTRW & ($\propto t^{\alpha}$,$\propto t^{\alpha}$) & $(>2,>1.49)$, eq.
(\ref{CTRW-MSD}) and (\ref{CTRW-MME}) & $\mathrm{Pr}\left(r\leq
r_0t^{\alpha/2},t\right) = A_0$\\
\hline
FBM & ($\propto t^{\alpha}$,$\propto t^{\alpha'}$), eq. (\ref{eq:FBM-MME})
& $(2,<1.49)$, eq. (\ref{eq:FBM-ratio}) & $\mathrm{Pr}\left(r\leq
r_0t^{\alpha/2},t\right) = A_0$
\end{tabular}
\caption{Test for 2D trajectories in a free environment, and equation
references for other dimensions. For each model, Brownian motion (BM),
diffusion on fractal, continuous time random walk (CTRW), and fractional
Brownian motion (FBM), the second column lists the scaling behavior of the
second regular and MME moments ($\langle r^2\rangle$ and $\langle
r_{\mathrm{max}}^2\rangle$); the third column shows the relative values of
the regular and MME ratio ($\langle X^4 \rangle / \langle X^2 \rangle
^2$);
and the fourth column contains the scaling laws of the probability, at time
$t$, to be in a sphere growing like $t^{\alpha/2}$.}
\label{sumtab}
\end{table*}

(1) Obtain the anomalous diffusion exponent $\alpha$ from power law fit to
MSD and second MME moment.
Different subdiffusion mechanisms can the be determined as follows:
(2) Diffusion on a fractal has regular and MME moment ratios, that depend on
both $\alpha$ and the fractal dimension $d_f$. The fractal dimension is
smaller than the embedding Euclidean dimension.
(3) CTRW subdiffusion has regular and MME moments that depend on the
anomalous diffusion exponent $\alpha$. The ratios are larger than the
corresponding Brownian quantities. The probability to be in a sphere
growing like $t^{\alpha/2}$ is constant.
(4) FBM has the same ratios for regular moments as Brownian motion. The MME
second moment exponent is greater than $\alpha$, and the MME ratio is
smaller than the Brownian one. The probability to be in a sphere growing
like $t^{\alpha/2}$ is constant.

\section{Discussion}
\label{practic}

We now turn to the question how experimental data can be analyzed by help of
the tools established above. In a typical experiment a small particle is
tracked by a microscope, the motion being projected onto the focal plane
(2D), to produce a time series $\mathbf{r}(t)=(x(t),y(t))$ of the particle
positions. Given a set of $N$ trajectories $\mathbf{r}_i(t)$, with $n_i$
steps in trajectory $i$, we first calculate the distances to the starting
point,
\begin{equation}
r_i(t)=\sqrt{\left[x_i(t)-x_i(0)\right]^2+\left[y_i(t)-y_i(0)\right]^2},
\end{equation}
in the 2D projection of the motion monitored in the experiment. The
propagator is not directly accessible in an experiment. However, division
of the number of trajectories being at $r$ for a given time $t$ in the 2D
projection, by the total number of trajectories of length $n_i\geq t$,
leads to a good estimate of $P(r,t)$. We can therefore transform all the
previous integrals defining the moments into discrete sums, and apply
above methods.

\subsection{Regular and MME moments}

In discrete form the $k$th order moments become
\begin{equation}
\langle r^k (t)\rangle\approx\frac{1}{\mathcal{N}(t)}\sum_{i=1}^{\mathcal{N}
(t)}r_i^k(t)
\end{equation}
and
\begin{equation}
\langle r_{\mathrm{max}}^k(t)\rangle\approx\frac{1}{\mathcal{N}t)}\sum_{i=1}
^{\mathcal{N}(t)} \left(\max_{0\leq t'\leq t}\Big\{r_i(t')\Big\}\right)^k,
\end{equation}
for regular and MME statistics, respectively. Here $\mathcal{N}(t)$ is the
number of trajectories that are at least $t$ steps long.

Note that the discrete MME moments defined here do not correspond exactly to
the theoretical definition provided before. In fact, we do not have access
to the whole trajectory, but only some sample points of it, with a given
time step between two consecutive frames. The real $r_{\mathrm{max}}$ may
be reached in between two frames, and therefore would not be observed.
However after sufficiently long time the difference between the discrete
estimate calculated here and the real value from the continuous trajectory
becomes sufficiently small.

\begin{figure}
\includegraphics[height=8.8cm,angle=270]{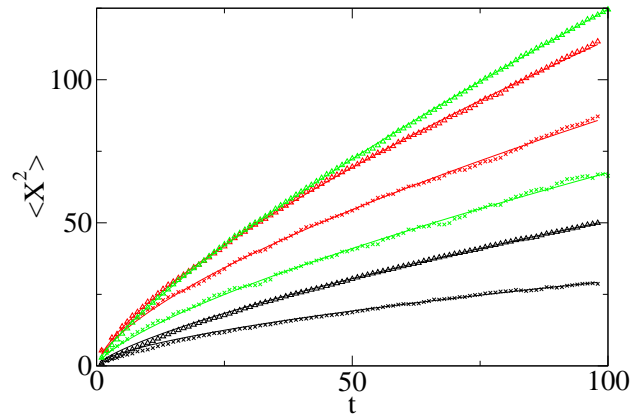}
\caption{MSD $\langle r^2(t)\rangle$ and second MME moment $\langle
r_{\mathrm{max}}^2\rangle$ as function of time $t$ (arbitrary units) for the
three simulated
time series ($1,000$ trajectories of $100$ steps each), each with
anomalous diffusion exponent $\alpha=0.7$. The power law fits produce: (i)
for 2D percolation data $\alpha=0.64$ (MSD, depicted by black $\times$)
and $\alpha=0.73$ (MME, black $\triangle$); (ii) for CTRW data
$\alpha=0.67$ (MSD, red $\times$) and $\alpha=0.71$ (MME, red
$\triangle$), (iii) for FBM data $\alpha=0.72$ (MSD, green $\times$) and
$\alpha'=0.79$ (MME, green $\triangle$, expected value
$\alpha'\approx0.74$).}
\label{test1}
\end{figure}

Figure \ref{test1} shows the result of fits of the MSD and the second MME
moment to simulated data according to the three subdiffusion models, all
with anomalous diffusion exponent $\alpha=0.7$. Indeed the MME method
performs somewhat better. We should note that these simulation results are
fairly smooth, and therefore we would not expect a significant difference
between the two methods, in contrast to the results on the experimental data
below. Also note that we chose different anomalous diffusion constants $K_{
\alpha}$ to be able to distinguish the different curves in figure
\ref{test1}.
Of course, this does not influence the quality of the fit of the anomalous
diffusion exponent $\alpha$.

Let us now turn to the moment ratios $\langle r^4\rangle/\langle
r^2\rangle^2$ and $\langle r_{\mathrm{max}}^4\rangle/\langle
r_{\mathrm{max}}^2\rangle^2$.
As mentioned above some care has to be taken with the latter: only the long
time values have a physical meaning. In fact, for the first frame, the
moment estimate $\langle r_{\mathrm{max}}^2\rangle$ is exactly $\langle
r^2\rangle$, because of the discrete time step. After few dozens of
frames, the estimate $\langle r_{\mathrm{max}}^2\rangle$ converges toward
its correct value, and the ratios become meaningful.

\begin{figure}
\includegraphics[height=8.8cm,angle=270]{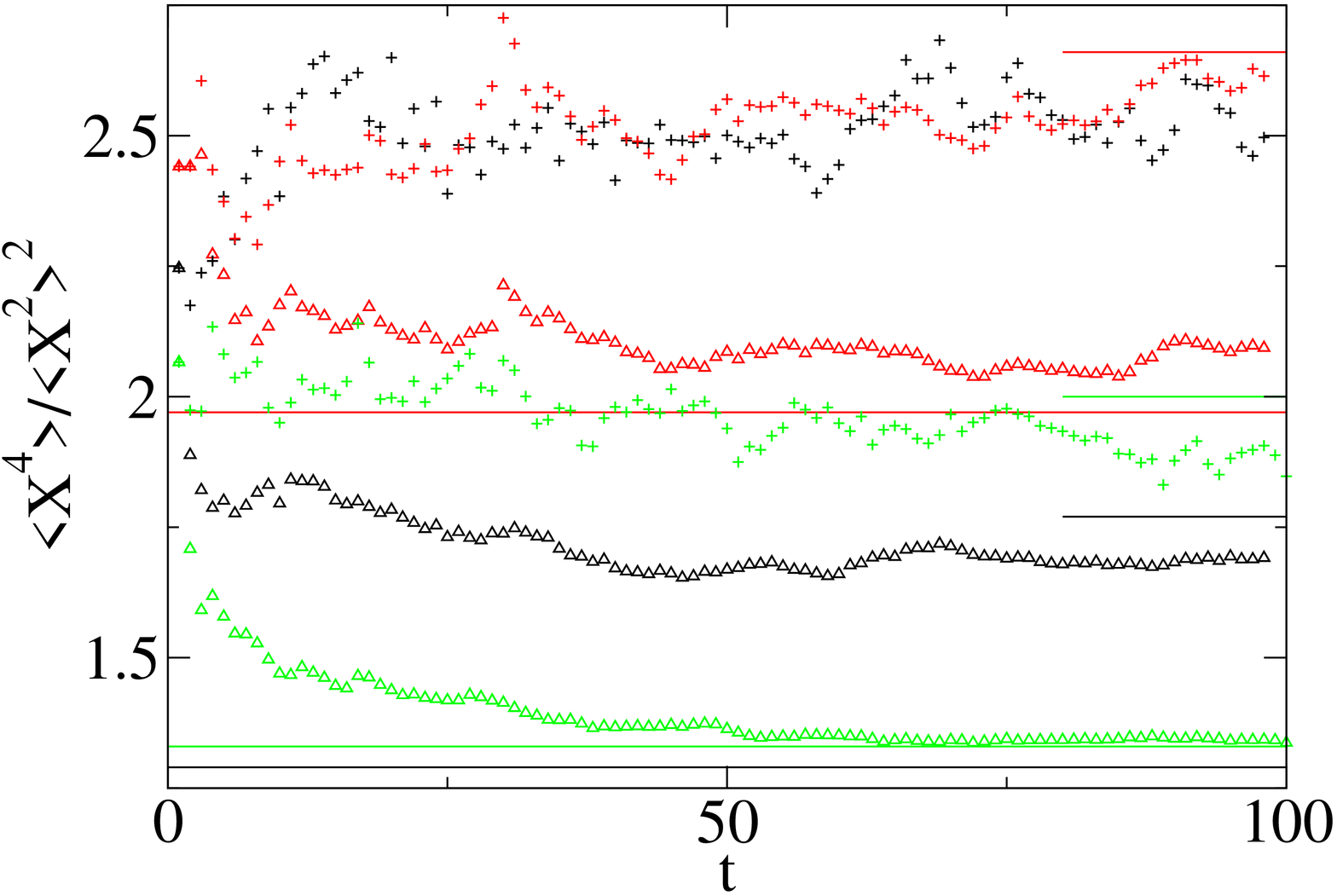}
\caption{Regular and MME moment ratios $\langle r^4\rangle/\langle r^2
\rangle^2$ and $\langle r_{\mathrm{max}}^4\rangle/\langle r_{\mathrm{max}}^2
\rangle^2$ as function of time (a.u.)
for the three simulated sets (diffusion on a
fractal, FBM, and CTRW). Each set consists of $1,000$ trajectories with
$100$ steps each.\\
Black $\triangle$: MME ratio for the diffusion on a 2D percolation
cluster; the data do not converge to the expected value 1.29 (black
horizontal line). The same behavior is observed for the regular moment
ratio (black $+$), for which the expected value is 1.77 (short black
line). This discrepancy is likely due to the confinement of the
percolation cluster on a $250\times250$ network: the random walker quickly
reaches the boundaries, and the convergence occurs toward the equilibrium
distribution, not toward the free space propagator.\\
Red $\triangle$: MME ratio for the CTRW process, converging to $1.97$ (red
horizontal line). We also plot the regular moment ratio (red $+$); these
are more irregular and converge to $2.66$ (short red line).\\
For FBM, the MME ratio (green $\triangle$) converges to the estimated
value of equation (\ref{eq:FBM-ratio}), $1.33$ (green horizontal line),
and the regular ratio (green $+$) oscillates around the Brownian value $2$
(short green line).}
\label{test3}
\end{figure}

In figure \ref{test3} we show a plot of the moment ratios. The convergence
to a constant value attained at sufficiently long times is distinct. The
ratios are those predicted for both CTRW and FBM, where the simulation is
performed in a free environment. For diffusion on a percolation cluster, we
observe a deviation from the prediction, due to the confinement of the
diffusion for this set: the propagator does not converge toward the free
space propagator, but toward the stationary distribution. We note that 
these ratios are clearly distinguishable between
regular and MME moments, but also between the three simulations sets.
Knowing the $\alpha$ value from the previous power law fit of MSD or
second MME moment, those ratios are already a good indication of the
underlying stochastic process. Since the difference between CTRW and
diffusion on a fractal is not too large, we use the method of a growing
sphere to see whether we can discriminate more clearly between those two
mechanisms.

\subsection{Growing sphere analysis}

\begin{figure}
\includegraphics[height=8.8cm,angle=270]{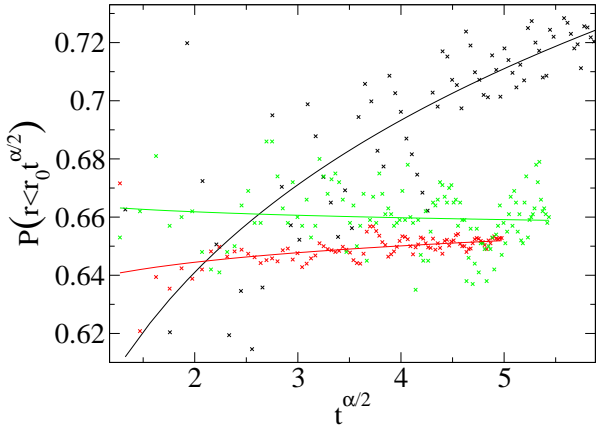}
\caption{Probability to be in a growing sphere of radius $r_0t^{\alpha/2}$ as
function of $t^{\alpha/2}$ for the three simulated sets (a.u.). This
analysis is based on the previously fitted values of $\alpha$. Results:
(i) 2D critical percolation (black $\times$) produces $d-d_f\approx0.11$,
i.e., $d_f\approx1.89$ (exact value $91/48\approx1.896$). (ii) The CTRW
set (red
$\times$) gives $d-d_f\approx0.01$ instead of $0$, and the FBM set (green
$\times$) leads to $d-d_f\approx-0.004$ instead of $0$.}
\label{test2}
\end{figure}

Let us turn to the probability to find the particle at time $t$ in a
(growing) sphere of radius $r_0 t^{\alpha/2}$. Here $r_0$ is a free
parameter. It should be chosen sufficiently large, such that for a given
trajectory the probability to be within the sphere is appreciably large. At
the same time it should not be too large, otherwise the probability to be
within the sphere is almost one.
Choosing a small multiple of $\langle r(t=1)\rangle$ appears to be a good
compromise. The probability to be inside the sphere then becomes
\begin{equation}
\mathrm{Pr}\left(r\leq r_0t^{\alpha/2}\right)\approx\frac{1}{\mathcal{N}(t)}
\sum_{i=1}^{\mathcal{N}(t)}\Theta\left(r_i(t)-r_0 t^{\alpha/2}\right).
\end{equation}
Here $\Theta(r)$ is the Heaviside function, that equals 1 if $r\ge0$, and 0
if $r<0$. We expect the scaling $\propto t^{\alpha(d-d_f)/2}$. To fit the
fractal dimension $d_f$ we need the anomalous diffusion exponent $\alpha$ as
input. We used the value extracted from the second MME moment fits. The
direct plot of the probability is quite easy to interpret: if the
probability is constant, then $d = d_f$; if it grows slowly,
then $d > d_f$, and the support is fractal ($d_f\neq d$). The dimension
$d$ here is the dimension of the trajectories ($d=2$ in our examples due
to the projection onto the focal plain).
In figure \ref{test2}, we see clearly that for CTRW and FBM the probability
is approximately constant, and that for the diffusion on a percolation
cluster, it grows with time, indicating that $d_f < d$, as it should be.

\subsection{Experimental data}

We analyse experimental single particle tracking data showing that such
time series are sufficiently large to apply the analysis tools
developed herein.

\begin{figure}
\includegraphics[height=8.8cm,angle=270]{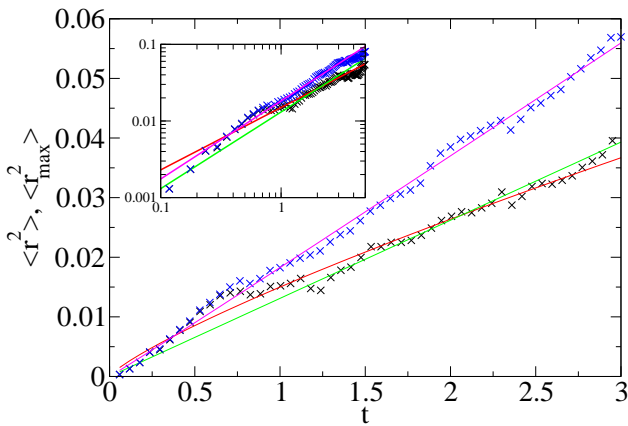}
\caption{Analysis of an experimental set of $67$ trajectories, the longest
consisting of $210$ points, for quantum dots freely diffusing in a solvent.
MSD (black $\times$), fitted by a power law with exponents $\alpha=0.81$ (red
line). We also show a fit with fixed exponent $\alpha=1$ (green line, expected
behavior for Brownian motion).
MME (blue $\times$), fitted by a power law (red line,
$\alpha=1.02$). Time is in seconds, distances are in $\mu$m$^2$.
Inset: double-logarithmic plot of the same data.}
\label{exp1:msd}
\end{figure}

The first data set (see supplementary material) contains 67 trajectories with
up to 210 steps length of quantum dots diffusing freely in a solvent. The
expected behavior is regular Brownian motion. The data set is quite small and
we show that MME moments are better observables than regular moments. We plot
the MSD as a function of time in figure \ref{exp1:msd}, and fit the data by a
power-law $\propto t^{\alpha}$. This fit provides an anomalous diffusion
coefficient $\alpha=0.81$.
The fit based on the second MME moment returns the
value $\alpha=1.02$, an almost perfect reproduction of the expected value
$\alpha=1$.
The much better result of the MME method is due to the
lower dispersion around the mean of the MME statistics, as discussed in the
supplementary material. In figure \ref{exp1:msd} it can be appreciated that
the large outlier in the MSD statistics at around $t=0.7$ sec is
responsible for the low $\alpha$ value. At longer times also the MSD follows
normal diffusion. This analysis demonstrates that the MSD in this case would
lead to a large deviation from the expected value, and thus to the erroneous
conclusion that the observed motion were subdiffusive, while the MME analysis
performs much more reliably.

\begin{figure}
\includegraphics[height=8.8cm,angle=270]{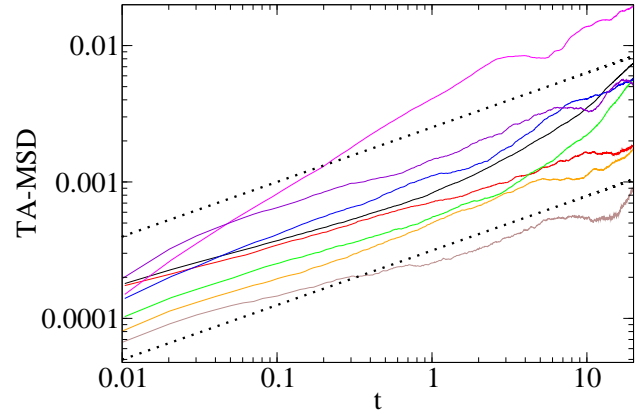}
\caption{Lipid granules diffusing in a yeast cell: $8$ trajectories,
between $5,515$ and $19,393$ frames long. Log-log plot of the
time-averaged MSD as a function of lag time (continuous lines), and $A_0
t^{0.4}$ (dotted lines). Time is scaled in seconds, and the time averaged
MSD in $\mu$m$^2$.}
\label{exp2:Dan}
\end{figure}

\begin{figure}
\includegraphics[height=8.8cm,angle=270]{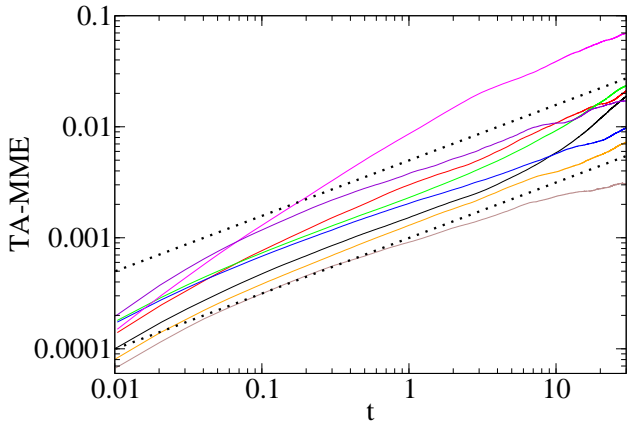}
\caption{Lipid granules diffusing in a yeast cell: log-log plot of the
time-averaged second MME moment of the data from figure \ref{exp2:Dan}
as function of lag time (continuous lines), and $A_0t^{0.5}$ (dotted
lines). Time is scaled in second, the ordinate is in $\mu$m$^2$.}
\label{exp2:DanMME}
\end{figure}

The second set of data was obtained from video tracking of 8 different
lipid granules moving in yeast cells. Since we had few long trajectories,
before an ensemble average, we first directly analyzed the 8
trajectories using the time-averaged MSD (\ref{tamsd}). We obtain a
distinct subdiffusive behavior with an exponent close to $0.4$, as
demonstrated in figure \ref{exp2:Dan}. Each trajectory corresponds to a
given granule.
It is interesting to see that the data exhibit a scatter in amplitude and
considerable local variation of slope. Such features were also observed
previously, see, for instance, references \cite{golding,lene}. They may
possibly be related to ageing effects \cite{stas}.
We also note that one of the curves shows a much steeper slope than the
others. We extended the time-average analysis to the second MME moment
\begin{equation}
\label{tamme}
\overline{\delta_{\textnormal{MME}}^2(\Delta,T)}=\frac{1}{T-\Delta}
\sum_{i=0}^{T-\Delta}\max_{i\leq t\leq i+\Delta}\Big\{r_i(t)\Big\}^2
\end{equation}
and again obtained a clear subdiffusive behavior, but with an exponent close
to 0.5, as demonstrated in figure \ref{exp2:DanMME}. Once again, we have a
scatter in amplitude. The initial slope variation ($0 < t < 10$) is due to
the inaccuracy in the MME estimation when there are only few frames to
average. A greater exponent for MME than for regular moment could be due
to an inaccuracy in the fit. However, it may indeed point toward an
underlying FBM process.

\begin{figure}
\includegraphics[height=8.8cm,angle=270]{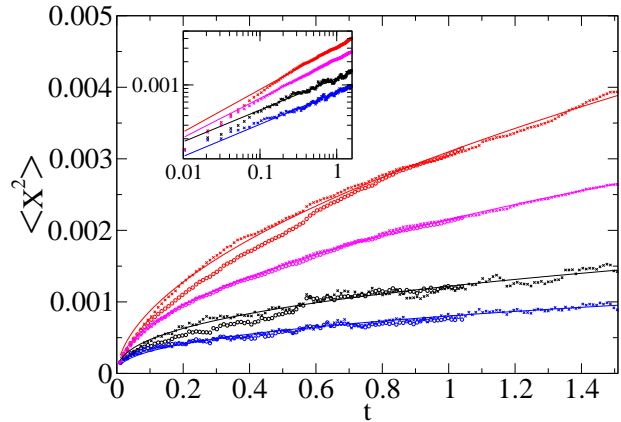}
\caption{Lipid granules diffusing in a yeast cell: $526$ subtrajectories
of $100$ steps extracted from the experimental set of $5$ trajectories,
that are between $5,515$ and $19,393$ frames long.
Ensemble averaged MSD (black $\circ$) fitted by a power law ($\alpha=0.41$,
black line), and ensemble averaged MME (red $\circ$), fitted with a power
law ($\alpha=0.55$, red line). We verified that creating $350$ trajectories
of $150$ steps instead of $100$ does not change the exponents obtained from
the MSD or the second MME moment ($\times$ instead of $\circ$ symbols).
Since one of the trajectories had a steeper slope than the others, we
repeated the same analysis without this trajectory, the new subset
containing $445$ trajectories of $100$ steps, or $296$ of $150$ steps (MSD
in blue leading to $\alpha=0.42$, second MME moment in magenta producing
$\alpha=0.51$). Time is in seconds,
the ordinate is measured in $\mu$m$^2$.}
\label{exp2:msd}
\end{figure}

In order to gain more insight into the diffusion mechanism producing this
subdiffusion behavior, we applied the methodology detailed above. Since the
different trajectories were not all recorded at the same frequency (96.5
and 99.1 frames per second), we kept only the greater set (96.5 fps),
containing 5 trajectories, and we split those into 526 short trajectories
of 100 steps each. These trajectories are non overlapping and one may view
them as the result of 526 separate observations. Surprisingly, we retrieve
the exponent $0.41\pm0.01$ using the MSD, and the value $0.53\pm 0.02$
from the second MME moment, as shown in figure \ref{exp2:msd}.
We repeated this analysis with a step size of 150 (350 trajectories)
concluding that the choice of the step size 100 has no influence on the
value of those coefficients. Since one of the trajectories (the magenta line
in figures \ref{exp2:Dan} and \ref{exp2:DanMME}) shows a much steeper slope, we
excluded it for the rest of the analysis.

An interesting observation is the following: assuming that the underlying
stochastic process is indeed an FBM, relation (\ref{eq:FBM-MME}) for
$\alpha=0.41$ predicts a value $\alpha'=0.50$ for the MME statistics, in
quite good agreement with the fitted value. This finding is quite suggestive
in favor of FBM as the stochastic process governing the particle motion.

Since the trajectories correspond to different granules, in different cells,
we also studied them separately: each trajectory was split into stretches
of 100 steps. For each granule, we plotted the regular and the MME ratios.
They are somewhat noisy, but for each granule the MME ratio is clearly
below the Brownian one (1.49): it ranges between 1.20 and 1.40. The
regular moment ratio is slightly above the Brownian value (2), between 1.7
and 2.5, as shown in figure 3 of the supplementary material. In the same
figure we also plotted the ratio for the whole set of
100 steps pieces (thick lines), which give approximately the same results
as those obtained for individual trajectories. From these ratios, we
obtain another clue pointing at an underlying FBM mechanism: the MME
moment ratio is, on average, below the value for Brownian motion, and the
regular moment ratio close to the Brownian value. These MME ratios are not
very precise, but seem to range somewhat above the expected value for FBM
with $\alpha=0.41$: equation (\ref{eq:FBM-ratio}) gives $1.21\pm0.02$.

The test with the growing sphere is, once again, somewhat noisy, however, it
clearly shows that the probability to be in a sphere, growing like
$t^{\alpha/2}$, attains a constant value (see figure 4 of the supplementary
material). This
excludes the possibility that the process corresponds to diffusion on a
fractal.

The above analysis demonstrates that the tools proposed in this study
allow us to classify the stochastic process underlying the motion of the
measured single particle trajectories of the granules. We observe that
this motion shares several distinct features with an FBM process. Namely
FBM explains the finding of different scaling exponents of the MSD and the
MME second moment, including their actual values connected by equation
(\ref{eq:FBM-MME}). It is also consistent with a Brownian regular moment
ratio, and an MME ratio lower than the Brownian one (compare figure 3 of
the supplementary material).
The recorded data were also shown to be incompatible with diffusion on a
fractal. So what about CTRW as potential mechanism? The scatter between
different single trajectories observed in the time averaged second moments
is reminiscent of the weak ergodicity breaking for CTRW subdiffusion with
diverging characteristic waiting time, as studied in references
\cite{he,lubelski}. However an alternative explanation may simply be
different environments and granule sizes. It should be noted that even
between successive recordings the cellular environment may change slightly,
influencing the motion of the observed particle. The CTRW hypothesis however
is not consistent with the moment ratio test: the expected ratio for $\alpha 
= 0.4$ would be 3.38 for the regular one, and 2.50 for the MME, far above 
the observed values.

Given the clues we obtained from the analysis, the experimental data quite
clearly point toward an FBM as underlying stochastic process. More
extensive data acquisition is expected to allow more precise conclusions.

\section{Conclusions}

With modern tracking tools biophysical experiments provide us with the time
series of single particle trajectories. Recently a growing number of cases
have been reported in which the monitored particles exhibit subdiffusion. An
important example is the motion of biopolymers under cellular crowding
conditions. While the mean squared displacement of these data, scaling like
$\simeq t^{\alpha}$, provides the anomalous diffusion exponent $\alpha$, the
underlying physical mechanism causing this subdiffusion is presently unknown.
As different mechanisms give rise to fundamentally different physical
behaviors influencing the particle diffusion in a living cell, it is
important to obtain information from experimental or simulation data other
than the anomalous diffusion exponent, allowing us to pin down the specific
stochastic process. We here introduced and studied several observables to
analyze more quantitatively single particle trajectories of
freely (sub)diffusing particles. For long trajectories with active motion
events the latter may be singled out and our analysis performed on the
passive parts of the trajectories \cite{raedler}. As typical experimental data
sets are relatively short, we here focus on the ensemble average obtained
from a larger number of individual trajectories. The data were simulated
on the basis of three subdiffusion models, these being continuous time
random walk with power law waiting time density, fractional Brownian
motion, and diffusion on a fractal support. Moreover we analyzed two
sets of experimental single particle tracking data, corresponding to a
Brownian and a subdiffusive system.

In particular we propose alternative measures to the usual fit to the
mean squared displacement. Apart from obtaining the fourth order moment
and construct the ratio $\langle r^4\rangle/\langle r^2\rangle^2$,
these alternatives are: (i) mean maximum excursion statistics that the
particle has not traveled more than a preset distance up to time $t$. Its
second and fourth moments, theoretically, scale with time the same way as
the regular moments; however, they appear to reproduce more truthfully the
actual subdiffusion exponents. Constructing the ratio $\langle
r_{\mathrm{max}}^4\rangle/\langle r_{\mathrm{max}}^2\rangle^2$ for these
quantities provides additional information, that allows one to distinguish
different subdiffusion mechanisms. (ii) The analysis using a growing sphere
containing a certain portion of particles appears as a quite reliable
method to obtain the (fractal) dimension of the underlying trajectory.

An application to an experimental set proves the efficiency of those tests:
the MME analysis is clearly more accurate than the classical MSD one, and
with a modest data set we are able to collect several independent clues
to identify FBM as mechanism to explain the motion of lipid granules under
molecular crowding conditions. For long recorded time
series the performance of the MME and regular moments analysis becomes
comparable.

From the discussion of simulations and experimental data is was shown that
in order to understand the physical mechanism of anomalous diffusion in a
given set of data one needs to gather evidence from complementary measures,
such as the ones proposed in this study.

\begin{acknowledgments}
We are grateful to Eli Barkai, Jae-Hyong Jeon, Yossi Klafter, and Igor
Sokolov for many helpful discussions.
We acknowledge funding from the Deutsche Forschungsgemeinschaft and the
CompInt graduate school at the Technical University of Munich.
\end{acknowledgments}

\end{document}